\documentstyle[12pt,epsfig]{article}

\def\be{\begin{equation}}
\def\ee{\end{equation}}

\begin{document}

\begin{titlepage}
\setlength{\textwidth}{5.0in}
\setlength{\textheight}{7.5in}
\setlength{\parskip}{0.0in}
\setlength{\baselineskip}{18.2pt}
\setlength{\footskip}{0.5in}
\setlength{\footheight}{0in}

\renewcommand{\thefootnote}{\fnsymbol{footnote}}


\vspace{0.3cm}

\begin{center}
{\Large\bf The warped product approach to magnetically charged
GMGHS spacetime}
\end{center}

\begin{center}
Jaedong Choi\footnote{Electronic address: choijdong@gmail.com}
\par

\end{center}

\begin{center}

{Aerospace Research Center, Korea Air Force Academy}\par
\end{center}

\vskip 0.5cm
\begin{center}
{\today}
\end{center}

\vfill

\begin{abstract}
In the framework of Lorentzian multiply warped products we study
the magnetically charged
Gibbons-Maeda-Garfinkle-Horowitz-Strominger (GMGHS) interior
spacetime in the string frame. We also investigate geodesic motion
in various hypersurfaces, and compare their solutions of geodesic
equations with the ones obtained in the Einstein frame.
\end{abstract}

\vskip20pt

\end{titlepage}

\newpage

\section{ Introduction}
\setcounter{equation}{0}
\renewcommand{\theequation}{\arabic{section}.\arabic{equation}}

The spherically symmetric static charged black holes in the
four-dimensional heterotic string theory has been extensively
investigated ever since the discovery of the solutions by Gibbons,
Maeda~\cite{gm} and by Garfinkle, Horowitz, Strominger~\cite{ghs}.
Null geodesics and hidden symmetries in the Sen black hole was
investigated by Hioki and Miyamoto~\cite{hm08}, and Gad~\cite{gad}
studied both geodesic equations and geodesic deviation of the GMGHS
black hole solution. Fernando~\cite{Fernando} investigated null
geodesic motions of the same solution both in the Einstein and
string frame. In a Lorentizan multiply warped product spacetime, by
exchanging timelike and spacelike coordinates, we have also studied
geodesic equations of the GMGHS interior spacetime in the Einstein
frame~\cite{choi04}.

Recently, we have obtained the (5 + 1)/(5 + 2) dimensional global
flat embeddings of the GMGHS spacetime~\cite{KCP} according to the
Einstein/string frames, respectively. In the Einstein frame where
the solution is obtained from the action in form of the
Einstein-Hilbert, we need the (5 + 1) dimensional embedding. The
result is similar to the embedding of the Schwarzschild
spacetime~\cite{Fronsdal}. On the other hand, in the string frame
where strings directly couple to the metric of
$e^{2\phi}g_{\mu\nu}$, the global flat embeddings of the GMGHS
spacetime need one more time dimension like the
Reinnser-Nordstr\"om spacetime~\cite{Deser,KPS}. Even though the
solutions in the two frames are known to be related by a conformal
transformation so that they are mathematically isomorphic to each
other~\cite{Faraoni}, there are differences in some of the
physical properties of the black hole solutions in the two
frames~\cite{Casadio} including the above mentioned different
global flat embeddings. This difference motivates us to study
geodesic equations of the GMGHS interior spacetime in the string
frame and compare with the ones obtained in the Einstein frame.

In this paper, we study the magnetically charged GMGHS interior
spacetime and investigate how different the Lorentzian multiply
warped products are in the sting frame from the Einstein one. We
also investigate the geodesic motion near hypersurfaces of this
interior spacetime in the string frame and compare their solutions
of geodesic equations with the ones obtained in the Einstein frame.
We shall use geometrized units, i.e., G = c = 1, for notational
convenience.

\section{Magnetically charged GMGHS black hole in the framework of warped products}
\setcounter{equation}{0}
\renewcommand{\theequation}{\arabic{section}.\arabic{equation}}

In the string frame, the GMGHS action is described by
\begin{equation}
 S=\int d^4x
 ~\sqrt{-g}e^{-2\phi}(R+4(\nabla\phi)^2-F_{\mu\nu}F^{\mu\nu}),
\end{equation}
where $R$ is the scalar curvature, $\phi$ is a dilation, and
$F_{\mu\nu}$ is the Maxwell's field strength. Through the conformal
transformation of $g^S_{\mu\nu}=e^{2\phi}g^E_{\mu\nu}$, it is
related to the Einstein frame action~\cite{gm,ghs}. The Einstein
metric does not change when we go from an electrically to a
magnetically charged black hole, but since the dilaton $\phi$
changes sign, the string metric does change. As a result, we get the
GMGHS solution of the Einstein field equations which represents the
geometry exterior to a spherically symmetric static charged black
hole.

In the Schwarzschild coordinates, the line element for the
magnetically charged GMGHS black hole  metric in the exterior region
($r>2m$) has the form as follows
\begin{equation}\label{rs}
ds^2=-\frac{\Bigl({1-\frac{2m}{r}}\Bigr)}{\Bigl(1-\frac{Q^2}{mr}\Bigl)}dt^2
     +\frac{dr^2}{\Bigl({1-\frac{2m}{r}}\Bigr)\Bigl(1-\frac{Q^2}{mr}\Bigl)}
     +r^2(d\theta^2+\sin^2\theta d\phi^2).
\end{equation}
Here, the parameters $m$ and $Q$ are mass and charge, respectively.
When $Q\rightarrow 0$, it is reduced to the Schwarzschild spacetime,
and  as like in the Schwarzschild spacetime, the magnetically
charged GMGHS solution has an event horizon at $r=2m$. The surface
area of the sphere of the magnetically charged GMGHS black hole,
defined by $\int d\theta d\phi\sqrt{g_{\theta\theta}g_{\phi\phi}}$,
is also the same as the Schwarzschild spacetime.

On the other hand, the line element for the magnetically charged
GMGHS metric for the proper interior region can be described by
\begin{equation}\label{inside}
ds^2=-\frac{dr^2}{\Bigl({\frac{2m}{r}-1}\Bigr)\Bigl(1-\frac{Q^2}{mr}\Bigl)}
      +\frac{\Bigl({\frac{2m}{r}}-1\Bigr)}{\Bigl(1-\frac{Q^2}{mr}\Bigl)}dt^2+r^2(d\theta^2+\sin^2\theta d\phi^2),
\end{equation}
where $r$ and $t$ are now new temporal and spacial variables,
respectively. \vskip20pt

A multiply warped product manifold, denoted by $M=(B\times
F_1\times...\times F_{n}, g)$, consists of the Riemannian base
manifold $(B, g_B)$ and fibers $(F_i,g_i)$ ($i=1,...,n$) associated
with the Lorentzian metric~\cite{choi00}. In particular, for the
specific case of $(B=R,~g_B=-d\mu^{2})$, the magnetically charged
GMGHS metric (\ref{inside}) can be rewritten as a multiply warped
products $(a, b)\times_{f_1}R\times_{f_2} S^2$ by making use of a
lapse function
\begin{equation}\label{lapse}
N^2=\Bigl({\frac{2m}{r}-1}\Bigr)\Bigl(1-\frac{Q^2}{mr}\Bigl).
\end{equation}
This lapse function is well defined in the region of
$\frac{Q^2}{m}<r<2m$ to rewrite it as a multiply warped products
spacetime by defining a new coordinate $\mu$ as
\begin{equation}\label{mudef}
\mu=\int^r_0~
dx\frac{x}{\sqrt{(2m-x)\Bigl(x-\frac{Q^2}{m}\Bigr)}}=F(r).
\end{equation}
Setting the integration constant zero as $r\rightarrow
\frac{Q^2}{m}$, we have
\begin{eqnarray}
\mu&=&
-\sqrt{(2m-r)\Bigl(r-\frac{Q^2}{m}\Bigr)}\nonumber\\
& &-\Bigl(m+\frac{Q^2}{2m}\Bigr)
\tan^{-1}\left(\frac{2m^2+Q^2-2mr}{2\sqrt{m(2m-r)(mr-Q^2)}}
\right) +\frac{\pi(2m^2+Q^2)}{4m}. \label{schlapse3} \nonumber\\
\end{eqnarray}
\begin{figure}[t!]
   \centering
   \epsfbox{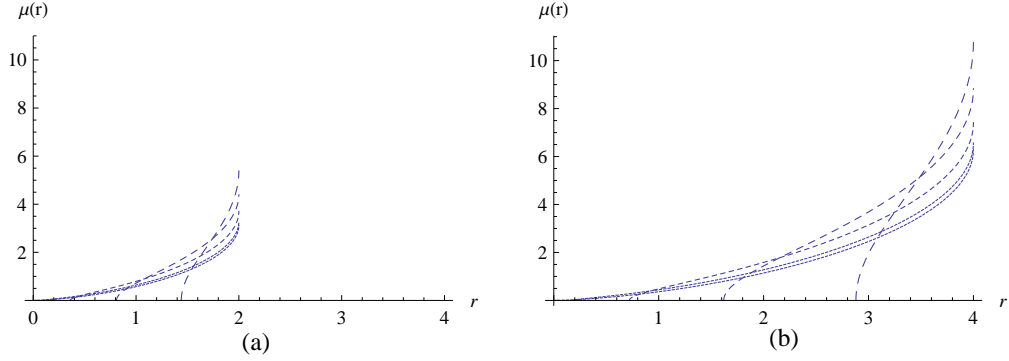}
\caption{$\mu$ as a function of $r$: Fig.~(a) is for $m=1$ and from
the top curve near $r=2$, $Q=1.2,~0.9,~0.6,~0.3$, while Fig.~(b) is
for $m=2$ and from the top curve near $r=4$, $Q=2.4,~1.8,~1.2,~0.6$.
The bottom curves denote the Schwarzschild limit of $Q=0$.}
\label{fig1}
\end{figure}
We have plotted $\mu$ as a function of $r$ in Fig.~\ref{fig1}. The
new coordinate $\mu(r)$ explicitly depends on the charge, while the
same coordinate in the Einstein frame~\cite{KCP} does not. We also
note that in the Einstein frame the GMGHS interior spacetime is
defined in $0<r<2m$ with the surface area singularity at
$r=\frac{Q^2}{m}$. However, the magnetically charged GMGHS interior
spacetime in the string frame is defined in $\frac{Q^2}{m}<r<2m$
with no surface area singularity. Thus, in Fig.~\ref{fig1}, we see
that as the charge increases, the proper range of $r$ is narrower.

Now, let us notice ${\frac{dF}{dr}}>0$ which implies $F^{-1}$ is a
well-defined function. We can thus rewrite the magnetically charged
GMGHS metric (\ref{inside}) with the lapse function (\ref{lapse})
and the two warping functions $f_1(\mu)$ and $f_2(\mu)$ as follows
\begin{eqnarray}
ds^2=-d\mu^2+f_1^2(\mu)dr^2+f_2^2(\mu)(d\theta^2+\sin^2\theta
d\phi^2),
\end{eqnarray}
where the warping functions are given by
\begin{equation}\label{fs}
f_{1}(\mu)=\left(\frac{\frac{2m}{F^{-1}(\mu)}-1}{1-\frac{Q^2}{F^{-1}(\mu)m}}
\right)^{1/2},~~~~ f_{2}(\mu)=F^{-1}(\mu), \label{schf1f2}
\end{equation}
and $F^{-1}(\mu)$ is defined in Eq.~(\ref{mudef}). We note that the
magnetically charged GMGHS interior spacetime written in the
multiply warped product spacetime has the same form with the
Kantowski-Sachs solution~\cite{KS}.

As a result, in the case of the interior region
$\frac{Q^2}{m}<r<2m$, the magnetically charged GMGHS metric has been
rewritten as a multiply warped product spacetime having the warping
functions in terms of $f_1(\mu)$ and $f_2(\mu)$ in Eq.~(\ref{fs}).
Moreover, we can write down the Ricci curvature on the multiply
warped products as
\begin{eqnarray}
R_{\mu\mu}&=&-\frac{f''_1}{f_1}-\frac{2f''_2}{f_2},\nonumber\\
R_{rr}&=&f_1 f''_1+\frac{2f_1f'_1f'_2}{f_2},\nonumber\\
R_{\theta\theta}&=&\frac{f'_1f_2f'_2}{f_1}+f'^2_2+f_2f''_2+1,\nonumber\\
R_{\phi\phi}&=&\left(\frac{f'_1f_2f'_2}{f_1}+f'^2_2+f_2f''_2+1\right)\sin^2\theta,\nonumber\\
R_{mn}&=&0,~{\rm for}~m\neq n,
\end{eqnarray}
which have the same form with the Ricci curvature of the multiply
warped interior Schwarzschild metric~\cite{choi00}. However, it is
differently parameterized due to the different functional
dependence as like in Eq.~(\ref{fs}).


\section{Geodesic motion near hypersurface}
\setcounter{equation}{0}
\renewcommand{\theequation}{\arabic{section}.\arabic{equation}}

In this section, we are interested in investigating the geodesic
curves of the magnetically charged GMGHS interior spacetime.
\vskip10pt

In local coordinates $\{x^i\}$  the line element corresponding to
this metric (\ref{inside}) will be denoted by
\begin{eqnarray}
dS^2 = g_{ij}dx^idx^j. \label{rf7}
\end{eqnarray}
Consider the equations of geodesics in the magnetically charged
GMGHS spacetime with affine parameter $\lambda$ given by
\begin{eqnarray}
\frac{dx^i}{d\lambda^2}+
\Gamma^i_{jk}\frac{dx^j}{d\lambda}\frac{dx^k}{d\lambda} = 0.
\label{rf8}
\end{eqnarray}

Let a geodesic $\gamma$ be given by $\gamma(\tau) = \Bigl(\mu(\tau),
r(\tau), \theta(\tau), \phi(\tau)\Bigr)$ of the magnetically charged
GMGHS interior spacetime of $\frac{Q^2}{m}<r<2m$ from
Eq.~(\ref{inside}), then the orbits of the geodesics equation are
given as follows
\begin{eqnarray} \frac{d^2\mu}{d\tau^2}+f_1{\frac{df_1}{d\mu}}\left(\frac{dr}{d\tau}\right)^2+f_2\frac{df_2}{d\mu}\left(\frac{d\theta}{d\tau}\right)^2+f_2\frac{df_2}{d\mu}\sin^2\theta\left(\frac{d\phi}{d\tau}\right)^2&=&0,\\
\frac{d^2r}{d\tau^2}+\frac{2}{f_1}\frac{df_1}{d\tau}\frac{dr}{d\tau}&=&0,\\
\frac{d^2\theta}{d\tau^2}+\frac{2}{f_2}\frac{df_2}{d\tau}\frac{d\theta}{d\tau}-\sin\theta\cos\theta\left(\frac{d\phi}{d\tau}\right)^2&=&0,\\
\frac{d^2\phi}{d\tau^2}+\frac{2}{f_2}\frac{df_2}{d\tau}\frac{d\phi}{d\tau}+{2\cot\theta}\frac{d\theta}{d\tau}\frac{d\phi}{d\tau}&=&0
\end{eqnarray}
with a following constraint along the geodesic
\begin{equation}
-\left(\frac{d\mu}{d\tau}\right)^2+f^2_1\left(\frac{dr}{d\tau}\right)^2+f^2_2\left(\frac{d\theta}{d\tau}\right)^2+f^2_2\sin^2\theta
\left(\frac{d\phi}{d\tau}\right)^2=\varepsilon.
\end{equation}
Note that a timelike (nulllike) geodesic is taken as
$\varepsilon=-1~(\varepsilon=0)$.
\vskip5pt
Hereafter, without loss of generality, suppose the geodesic
\begin{equation}
\gamma(\tau_0 ) = \Bigl(\mu(\tau_0 ), r(\tau_0 ), \theta(\tau_0),
\phi(\tau_0)\Bigr)
\end{equation}
for some $\tau_0$ and the equatorial plane of
$\theta=\frac{\pi}{2}$, thus $\frac{d\theta}{d\tau}=0$. Then, the
geodesic equations are reduced to
\begin{eqnarray} \label{mueq}\frac{d^2\mu}{d\tau^2}+f_1\frac{df_1}{d\mu}\left(\frac{dr}{d\tau}\right)^2+f_2\frac{df_2}{d\mu}\left(\frac{d\phi}{d\tau}\right)^2&=&0,\\
\label{req}\frac{d^2r}{d\tau^2}+\frac{2}{f_1}\frac{df_1}{d\tau}\frac{dr}{d\tau}&=&0,\\
\frac{d^2\theta}{d\tau^2}&=&0,\\
\label{phieq}\frac{d^2\phi}{d\tau^2}+\frac{2}{f_2}\frac{df_2}{d\tau}\frac{d\phi}{d\tau}&=&0
\end{eqnarray}
with a constraint
\begin{equation}\label{constrainteq}
-\left(\frac{d\mu}{d\tau}\right)^2+f^2_1\left(\frac{dr}{d\tau}\right)^2+f^2_2\left(\frac{d\phi}{d\tau}\right)^2=\varepsilon.
\end{equation}
These geodesic equations can be further simplified to give
\begin{eqnarray}
\label{req11}\frac{dr}{d\tau}&=&\frac{c_1}{f_1^2},\\
\label{phieq1}\frac{d\phi}{d\tau}&=&\frac{c_2}{f_2^2},\\
\frac{d^2\theta}{d\tau^2}&=&0\\
\end{eqnarray}
with a constraint
\begin{eqnarray}
\label{constrainteq11}-\left(\frac{d\mu}{d\tau}\right)^2+\frac{c_1^2}{f_1^2}+\frac{c_2^2}{f_2^2}&=&\varepsilon.
\end{eqnarray}
The constant $c_1$ represents the total energy per unit rest mass of
a particle as measured by a static observer~\cite{gad,Clarke,Wald},
and $c_2$ represents the angular momentum in the GMGHS spacetimes.
The equations for $r$ and $\phi$ are obtained from Eqs.~(\ref{req})
and (\ref{phieq}), respectively. Making use of these $r$, $\phi$
equations, we can show that Eq.~(\ref{mueq}) is the exactly same
with Eq.~(\ref{constrainteq}) when we take the integration constant
as $-\frac{\varepsilon}{2}$. \vskip5pt

\begin{figure}[t!]
   \centering
   \epsfbox{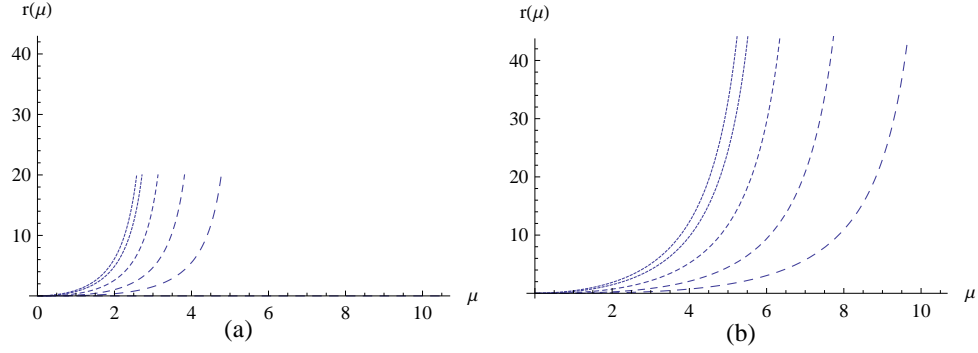}
\caption{Geodesic curves on the hypersurface $\Sigma_{r}$:
Fig.~(a) is for $m=1$ and, from the left curve, $Q=0$, the
Schwarzschild limit, and $Q=0.3,~0.6,~0.8,~1.2$, while Fig.~(b) is
for $m=2$ and, from the left curve, $Q=0$, the Schwarzschild
limit, and $Q=0.6,~1.2,~1.8,~2.4$. } \label{fig2}
\end{figure}

Now, we consider the null geodesics in the $r$-direction, which is
defined by the hypersurface $\Sigma_{r}$ by taking
$d\theta=d\phi=0$. Then, we have $c_2=0$ in Eq.~(\ref{phieq1}). Two
equations (\ref{req11}) and (\ref{constrainteq11}) are now reduced
to give
\begin{equation}\label{rgeod1}
dr=\frac{d\mu}{f_1(\mu)}.
\end{equation}
We have plotted in Fig.~\ref{fig2} the null geodesic curve on the
hypersurface $\Sigma_{r}$. We see the radial coordinate $r(\mu)$ is
a monotonic function of $\mu$. The curves on the far left in
Fig.~\ref{fig2} (a), (b) correspond to the Schwarzschild limit.
Comparing with the curve, Fig.~1 in Ref.~\cite{choi04}, obtained in
the Einstein metric, which has no charge dependence, we have
explicit charge dependence on $r(\mu)$. \vskip5pt

\begin{figure}[t!]
   \centering
   \epsfbox{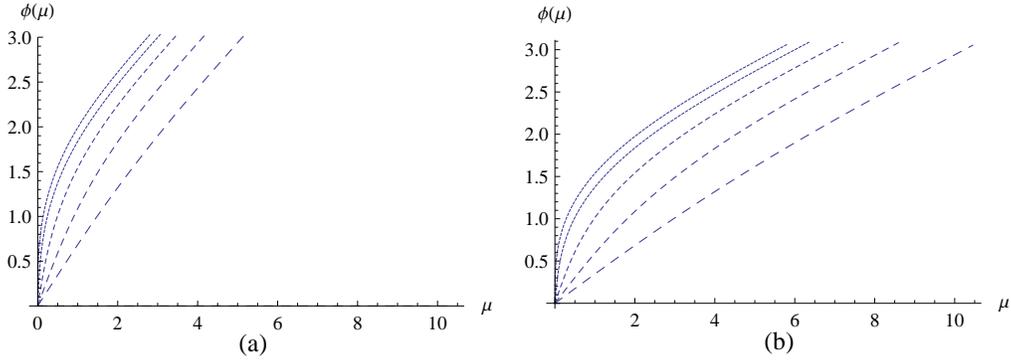}
\caption{Geodesic curves on the hypersurface $\Sigma_{\phi}$ at
the plane $\theta=\frac{\pi}{2}$ with $dt=0$: Fig.~(a) is for
$m=1$ and, from the top curve, $Q=0$, the Schwarzschild limit, and
$Q=0.3,~0.6,~0.8,~1.2$, while Fig.~(b) is for $m=2$ and, from the
top curve, $Q=0$, the Schwarzschild limit, and
$Q=0.6,~1.2,~1.8,~2.4$. } \label{fig3}
\end{figure}

Let us consider the geodesic in the $\phi$-direction, which lies
on the hypersurface $\Sigma_{\phi}$ at $\theta=\frac{\pi}{2}$ with
$dr=0$.  Then, we have $c_1=0$ in Eq.~(\ref{req11}). Two equations
(\ref{phieq1}) and (\ref{constrainteq11}) are reduced to give
\begin{equation}\label{phiDeq2}
d\phi=\frac{d\mu}{f_2(\mu)},
\end{equation}
where $f_2(\mu)$ is given by Eq.~(\ref{fs}). In Fig.~\ref{fig3}, we
have numerically drawn the azimuth angle. The top curve in
Fig.~\ref{fig3} corresponds to the zero charged pure Schwarzschild
limits. We note all the other curves depending on the charges start
from the origin. This contrasts with the corresponding curves in
Ref.~\cite{choi04} obtained from the geodesic in the
$\phi$-direction in the Einstein frame, where the curves start from
the points determined by the charges. \vskip5pt

\begin{figure}[t!]
   \centering
   \epsfbox{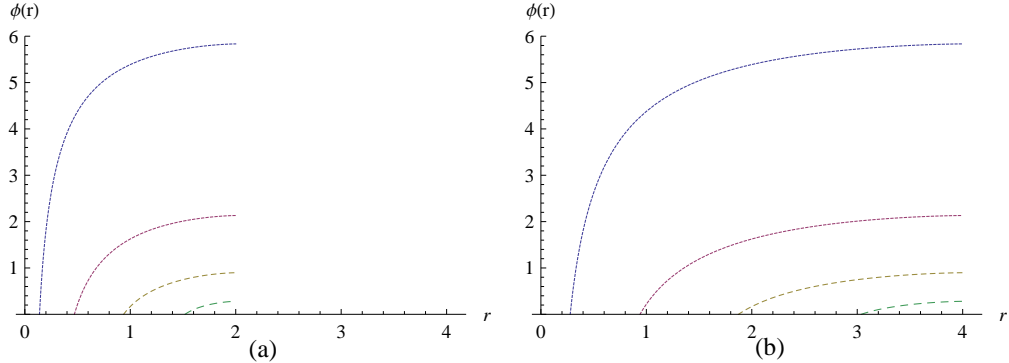}
\caption{Geodesic curves on the hypersurface $\Sigma_{\mu}$:
Fig.~(a) is for $m=1$ and, from the top curve,
$Q=0.3,~0.6,~0.8,~1.2$, while Fig.~(b) is for $m=2$ and, from the
top curve, $Q=0.6,~1.2,~1.8,~2.4$. } \label{fig4}
\end{figure}

Finally, let us find the geodesic in the $\mu$-direction, which is
defined by the hypersurface $\Sigma_{\mu}$, eliminating $\mu$ in
Eqs. (\ref{rgeod1}) and (\ref{phiDeq2}), leading to
\begin{equation}
\frac{d\phi}{dr}=\frac{1}{r}\sqrt{\frac{2m-r}{r-\frac{Q^2}{m}}},
\end{equation}
which is the exactly same as the one obtained in
Ref.~\cite{choi04}. In Fig.~\ref{fig4}, we have drawn the geodesic
curve $\phi(r)$ for different masses and charges. \vskip10pt

\section{Conclusions}
\setcounter{equation}{0}
\renewcommand{\theequation}{\arabic{section}.\arabic{equation}}
\label{sec:conclusions}
In this paper, we have studied the magnetically charged GMGHS
interior spacetime in associated with a multiply warped product
manifold. In the multiply warped product manifold, the magnetically
charged GMGHS spacetime has been characterized by two warping
functions $f_1(\mu)$ and $f_2(\mu)$, compared with the Schwarzschild
spacetime which has one warping function of $f_1(\mu)$ and the GMGHS
interior spacetime in the Einstein frame which have two $f_1(\mu)$,
$f_2(\mu)$ but different warping functions. We have also
investigated the geodesic motion near hypersurfaces in the interior
of the event horizon. Due to the charge term on the lapse function
(\ref{lapse}), the two warping functions are also charge dependent,
and as a result, the geodesic curves in the magnetically charged
GMGHS interior spacetime have been drawn by charges. This contrast
with the geodesic curves in Ref.~\cite{KCP} where the lapse function
has no charge dependent term.

\section*{Acknowledgement}

J. Choi would like to acknowledge financial support from Korea Air
Force Academy Grant (KAFA 14-02).

\end{document}